# The Baryonic Resonances
## with the Strangeness S=+1 in the System of nK+
## from the Reaction np®npK+K-
## at the Momentum of Incident Neutrons Pn=5.20±0.12GeV/c

**Yu.A.Troyan\*, A.V.Beljaev, A.Yu.Troyan, E.B.Plekhanov, A.P.Jerusalimov,G.B.Piskaleva,**
**S.G.Arakelian**

*E-mail: atroyan@jinr.ru

The investigation has been performed at the Veksler and Baldin Laboratory of High Energies, JINR.

These results were partially presented at the 32nd International Conference on High Energy Physics, Beijing, China, 16-22 August 2004

Dubna, 29 September, 2004

http://arxiv.org./hep-ex/0404003

D.Diakonov, V.Petrov and M.Polyakov have suggested in the papers [1,2] the development of the sheme of $SU(3)$-symmetry for states with the strangeness $S = +1$. It was claimed the existence of anti-decuplet $\overline{10}$, which included states consisting of 5 quarks ($uudd\overline{s}$). The dynamics of new resonances was based on the model of chiral soliton. This fact gave the possibility to estimate masses, widths and quantum numbers of expected new effects, to suggest the formula of the rotational band that gave a dependence of the resonance masses on their spins.

In the paper [1], $\Theta$-resonance at the mass $M = 1.530 GeV/c^2$, width $\Gamma \pounds 15 MeV/c^2$ and with quantum numbers $Y = 2$, $I = 0$, $J^P = 1/2^+$ is in the vertex of anti-decuplet.

The properties of the particles from anti-decuplet predicted in [1, 2] are the following that allow the direct search of effects. These are both comparatively low masses and accessible for a direct measurement widths. That is why there are a number of experimental works.

In present work we have attempted to study the characteristics of the predicted effects more detailed.

The study was carried out using the data obtained in an exposure of 1-$m$ HBC of LHE (JINR) to a quasimonochromatic neutron beam with $DP_n / P_n » 2.5\%, DW_n » 10^{-7} sterad$.

due to the acceleration of deuterons by synchrophasotron of LHE. [3]

The accuracy of the momenta of secondary charged particles from the reaction $np ® npK^+K^-$ are:

$dP » 2\%$ for protons and $dP » 3\%$ for $K^+$ and $K^-$.

The angular accuracy was $£ 0.5°$.

The channels of the reactions were separated by the standard $c^2$–method taking into account the corresponding coupling equations [4].
There is only one coupling equation for the parameters of the reaction $np ® npK^+K^-$ (energy conservation law) and the experimental $c^2$–distribution must be the same as the theoretical $c^2$–distribution with one degree of freedom.

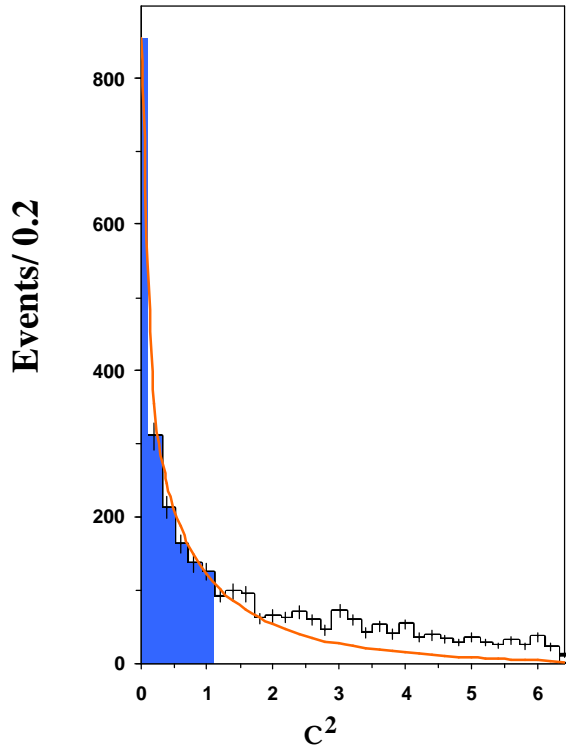

**np ⊗ npK⁺K⁻  Pₙ = 5.20 GeV/c**

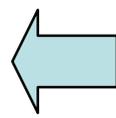
The experimental (histogram) and the theoretical (curve) $c^2$–distributions for the reaction *np ⊗ npK⁺K⁻*.

One can see a good agreement between them up to $c^2 = 1$

and some difference for $c^2 > 1$.

The missing mass distribution for the events of $c^2 £ 1$.

The distribution has the maximum at the value equal to the neutron mass with accuracy of 0.1 *MeV/c²*,

the width at the half-height $8 MeV/c^2$ and is symmetric about the neutron mass.

Later on a small number of events the missing masses
out of range pointed by arrows were excluded
for more purity of data.

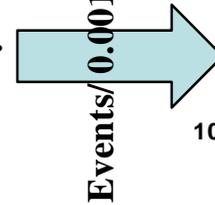

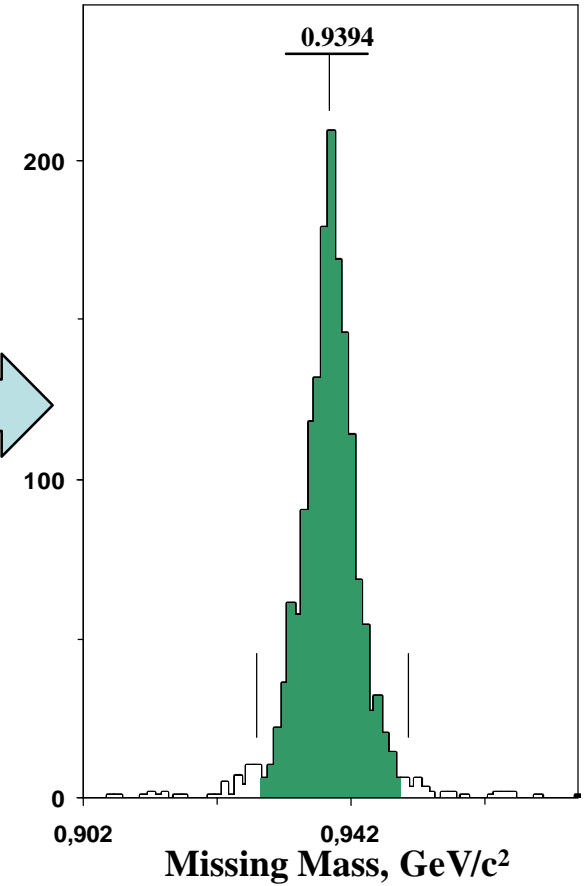

**np ⊗ npK⁺K⁻  Pₙ = 5.20 GeV/c**

The mass distributions is approximated by an incoherent sum
of the background curve taken in the form of a superposition
of Legendre polynomials and by resonance curves
taken in the Breight-Wigner form.

The requirements to the background curve are the following:
firstly, the errors of the coefficients must be not more than 50 %
for each term of the polynomial;
secondly, the polynomial must describe the experimental distribution
after "deletion" of resonance regions with $\overline{c^2} = 1.0$ and $\sqrt{D} = 1.41$
(the parameters of $c^2$–distribution with 1 degree of freedom).

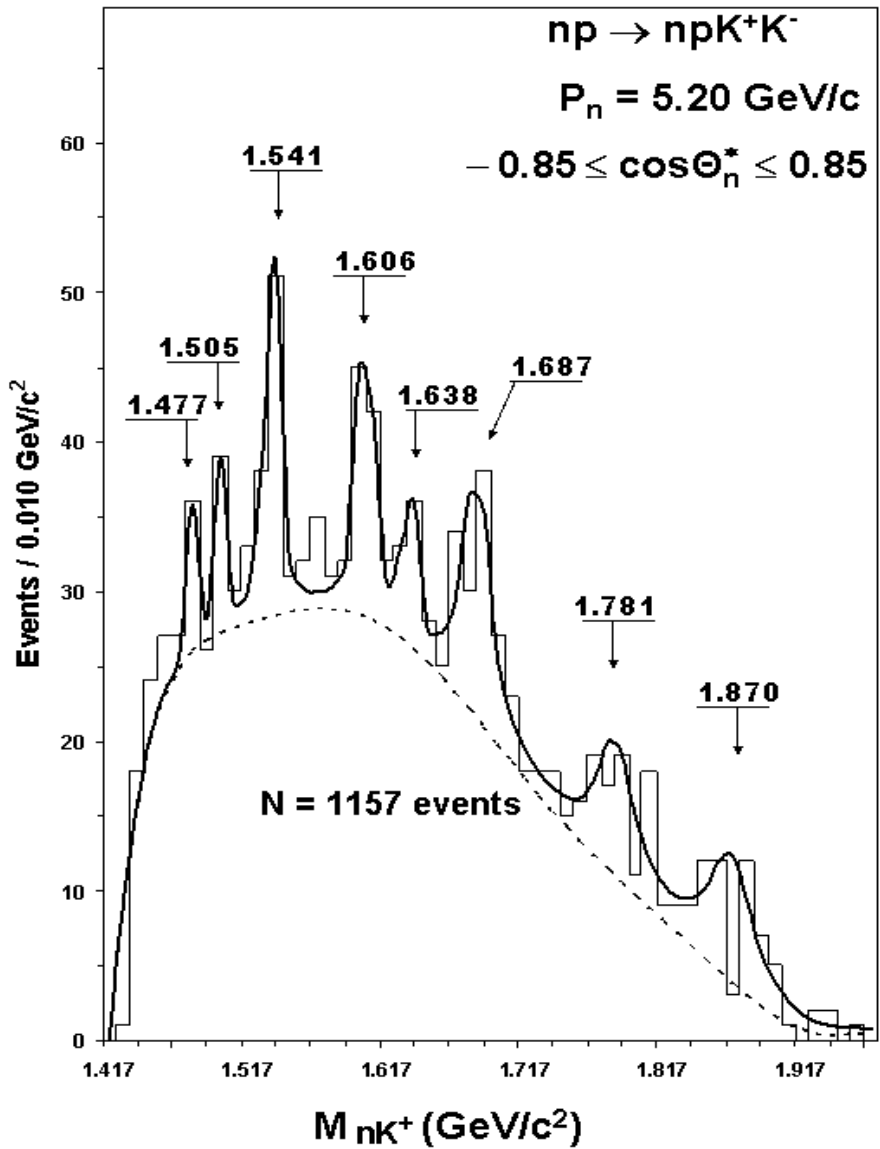

The distribution of effective masses of $nK^+$–combinations for the events selected under condition $-0,85 < \cos\Theta_n^* < 0,85$. This condition can decrease the background.

The parameters for the distribution in figure are $\overline{c^2} = 0.82 \pm 0.26$ and $\sqrt{D} = 1.40 \pm 0.19$.

The same parameters for the background curve normalized to 100 % of events (resonance regions are included) are $\overline{c^2} = 1.64 \pm 0.20$ and $\sqrt{D} = 2.90 \pm 0.14$.

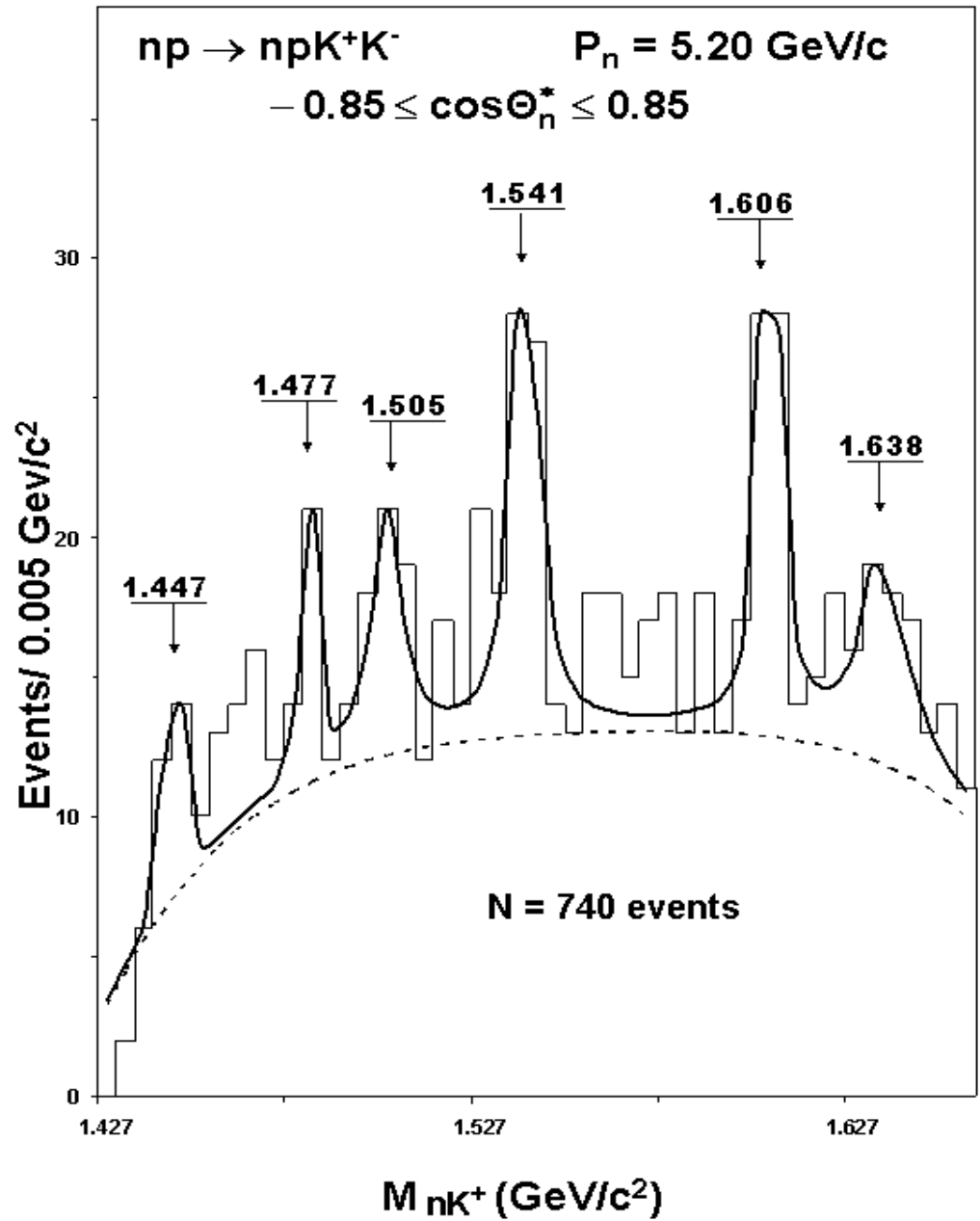

For a better study
of left part of the figure,
the distribution of effective masses
of $nK^+$ –combinations
was constructed
with bins of 5 $MeV/c^2$
(up to the mass of ~1.663 $GeV/c^2$)

Also, the widths of the resonances
are more precise determined
by means of this distribution

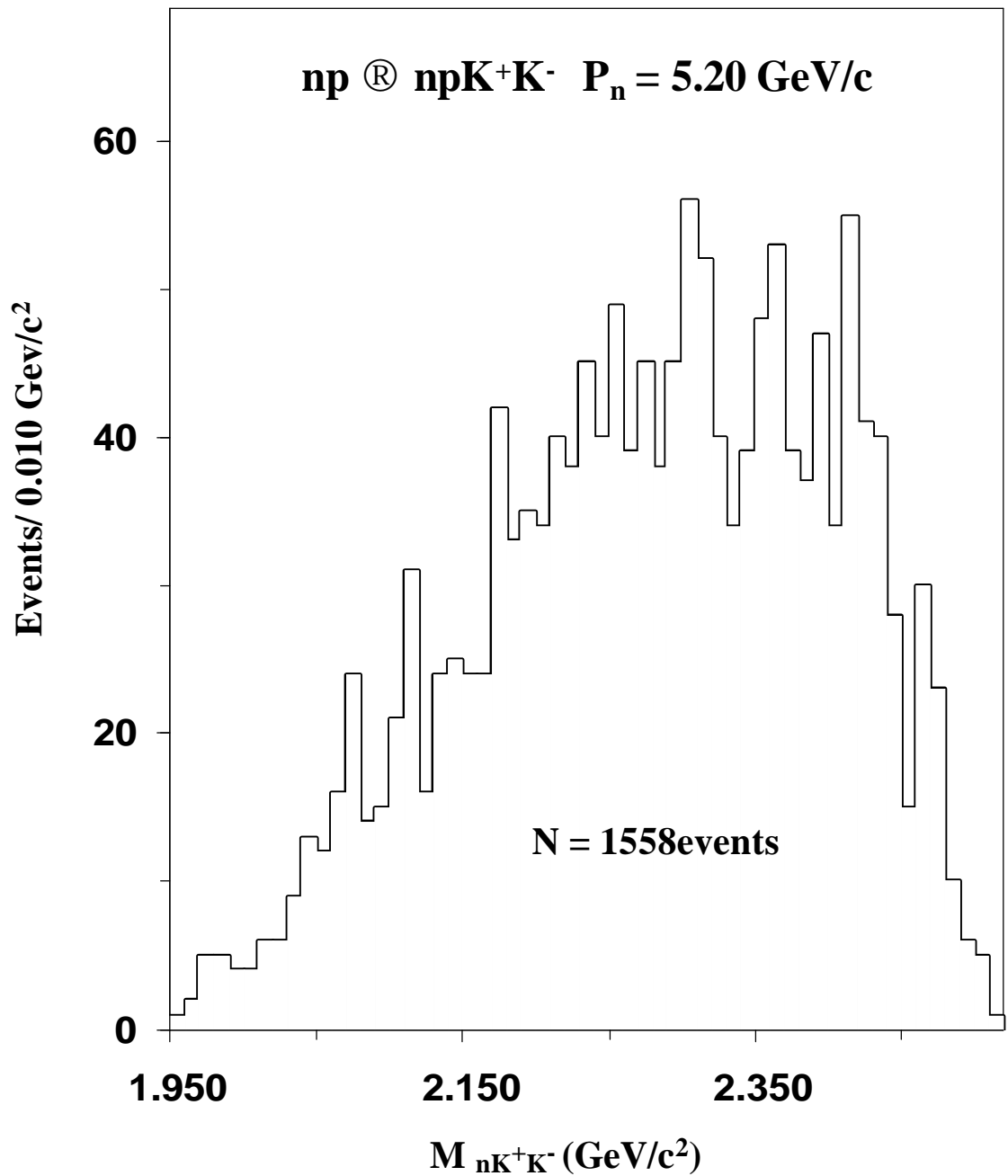

np ® npK⁺K⁻  $P_n = 5.20$ GeV/c

N = 1558events

**M** $_{nK^+K^-}$ **(GeV/c²)**

We have undertook an attempt to increase the statistical significances of some resonances. This attempt was based under the assumption that resonances were produced by means of *K*-exchange mechanism. The kinematically produced peaks could form in the effective mass distribution of $nK^+K^-$ -system. A number of peculiarities are clearly observed in this distribution. Corresponding resonances decaying through the mode $R \to NK\bar{K}$ are absent in PDG tables. These are just the same kinematic reflections.

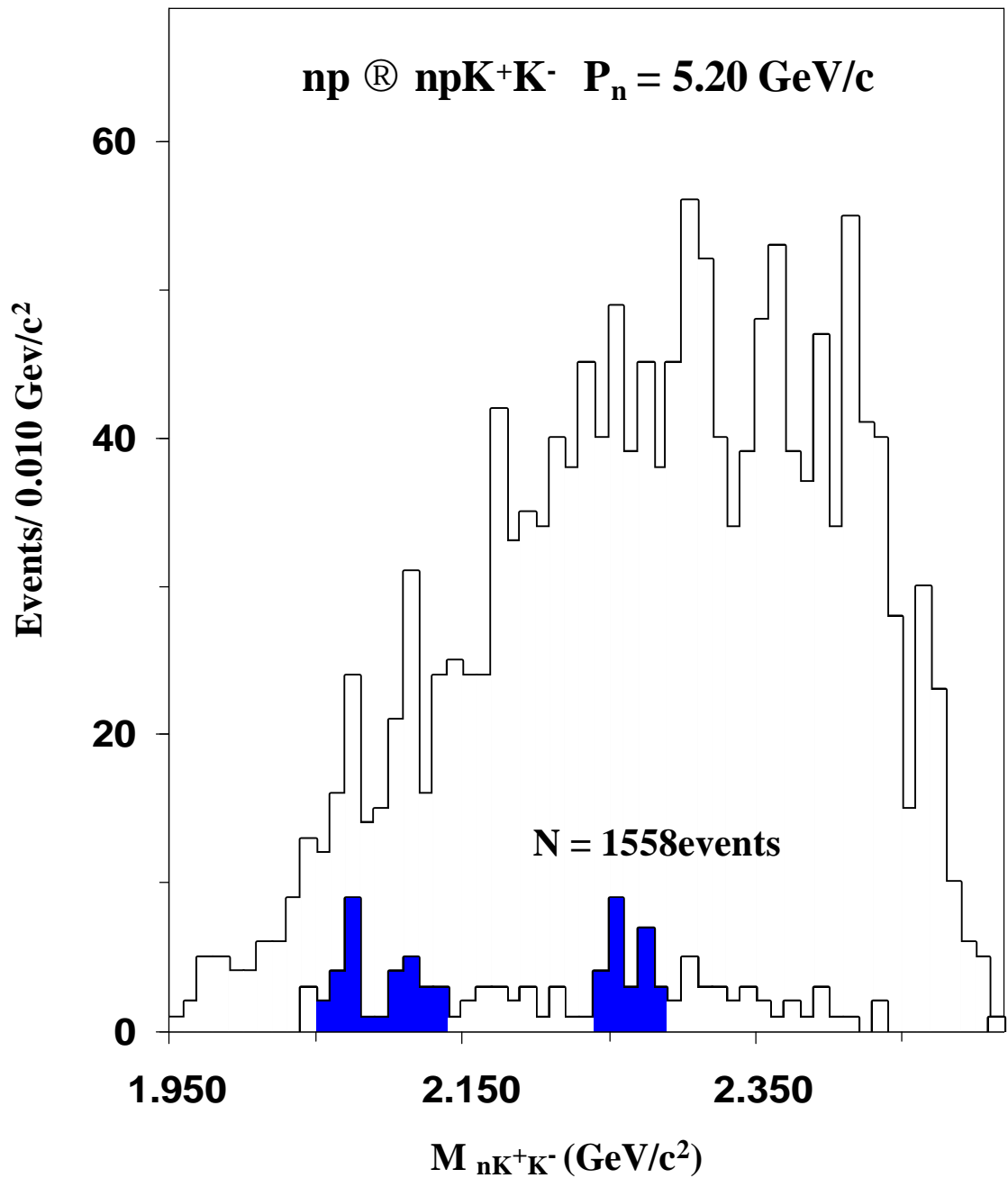

$np \rightarrow npK^+K^-$  $P_n = 5.20$ GeV/c

N = 1558 events

The same figure shows the effective mass distribution of $nK^+K^-$-combinations constructed under condition that the effective mass $nK^+$-system is within the range of the resonance at $M = 1.541 GeV/c^2$.

Two clusters are clear seen in this distribution (marked bins)

Corresponding clusters exist for resonances in $nK^+$-system at the masses of $M = 1.606 GeV/c^2$ and $M = 1.687 GeV/c^2$.

**Selecting regions of masses of $nK^+K^-$ -combinations that correspond to the $nK^+$ -resonances, we obtain the following distributions of effective masses:** for the resonances at the

$M = 1.541 GeV/c^2$,      $M = 1.606 GeV/c^2$,        $M = 1.687 GeV/c^2$

**We get a significant enhancement of effects :**

6.8 S.D.            5.2 S.D.              6.8 S.D.

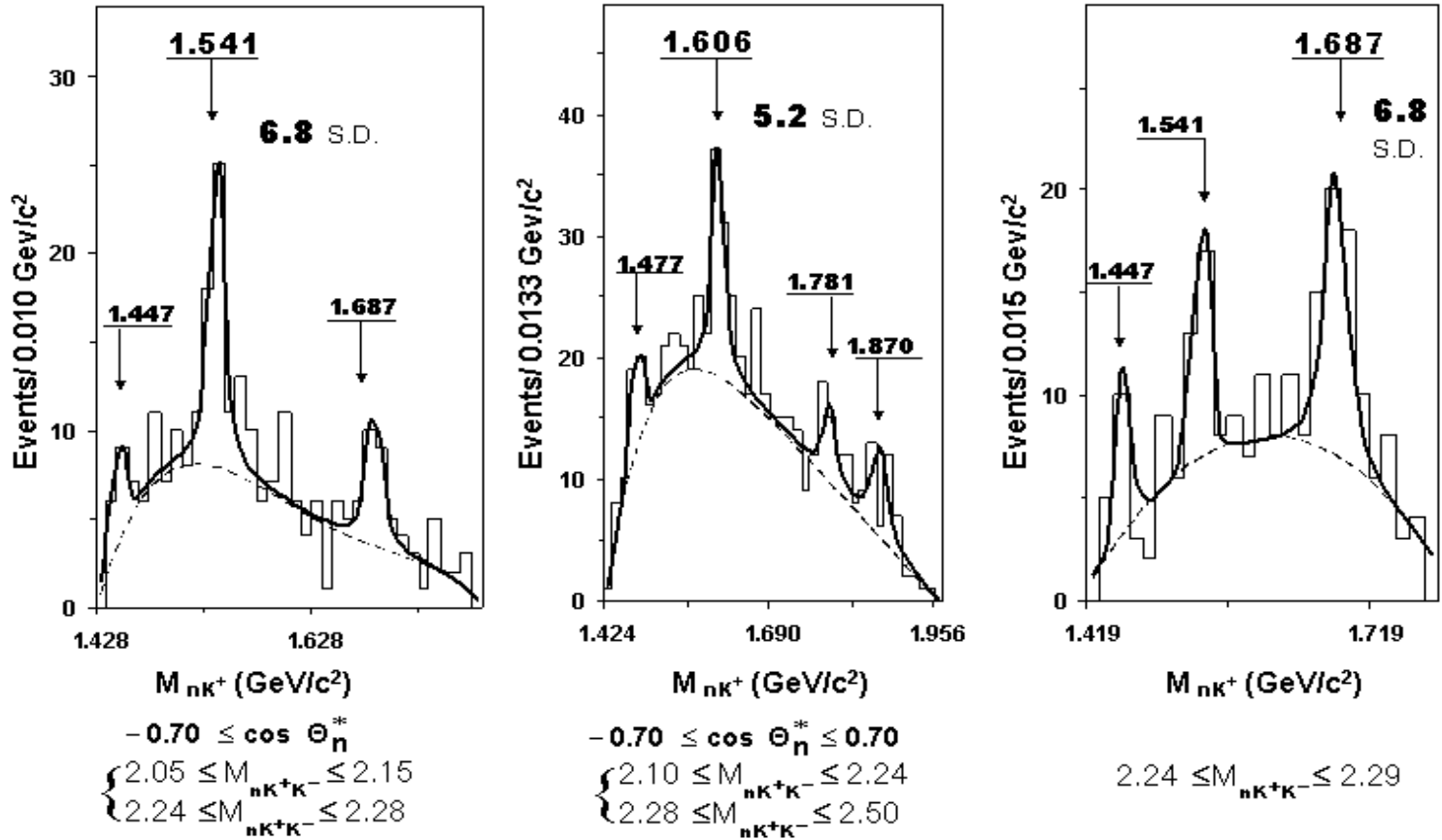

**In this case the number of events in peaks do not decrease as compared with the data presented in previous figures.**

To do this there were constructed the distributions of emission angles of neutron from reso nance decays in helicity coordinate system. In the helicity coordinate system, the angular distributions have to be described by sum of Legendre polynomials of even degrees and a maximum degree has to be equal to $(2J - 1)$ where J – spin of a resonance (for half-integer spins).

The backgrounds are constructed using events at the left and at the right of the corresponding resonance band and subtracted using the weight in proportion to a contribution of a background into resonance region.

By this manner the value of low limit of resonance spin was estimated.

**We have tried to estimate the values of spins for the observed resonances in $nK^+$-system.**

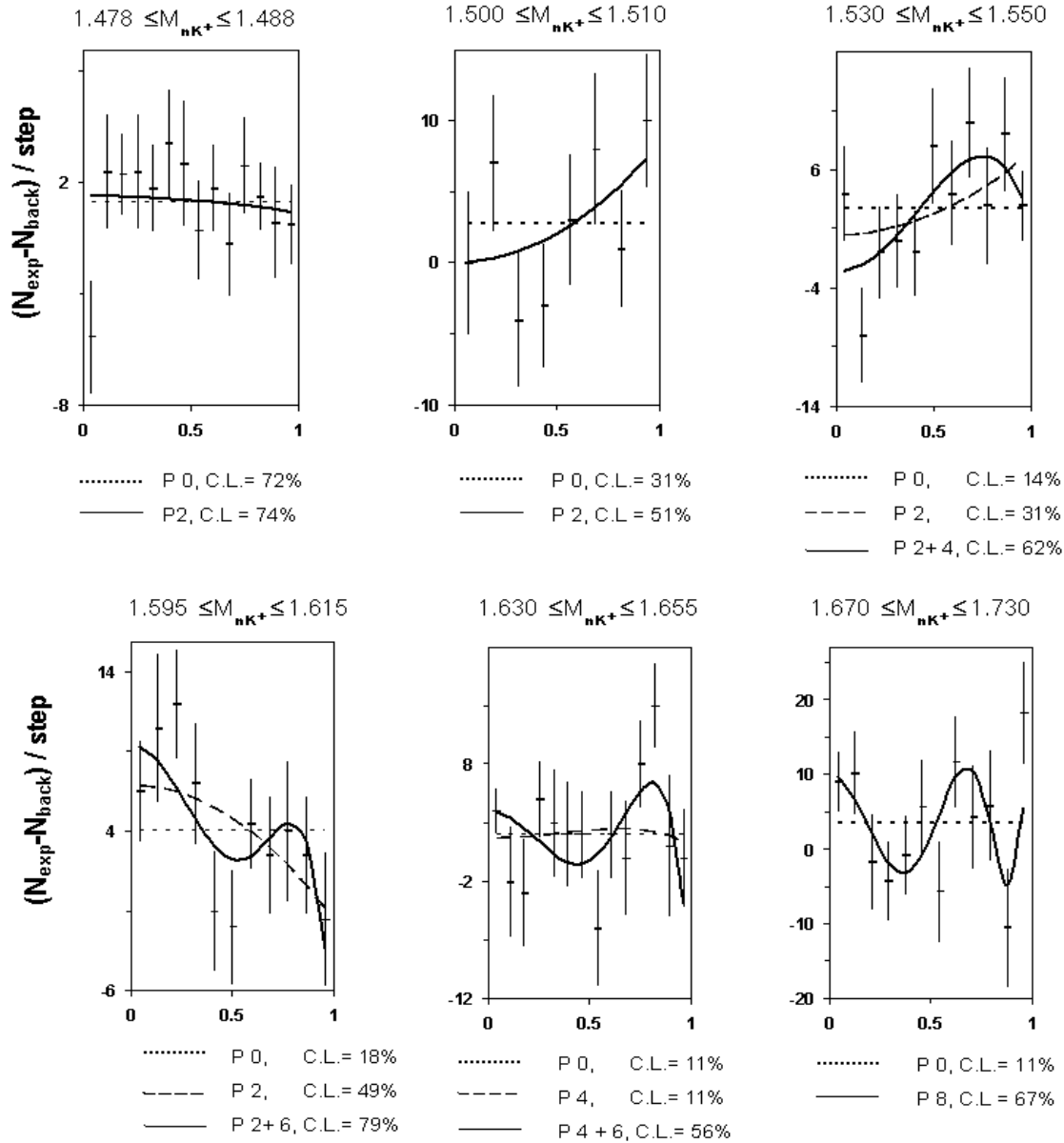

$np \to npK^+K^-$     $P_n = 5.20$ GeV/c

Figure shows
the angular distributions
for six resonances
which masses are within the ranges
pointed on plots
with confidence level for each
approximation

The value of spin $J \geq 1/2$
for the resonance
at the mass of $M = 1.541 GeV/c^2$
has a confidence level
significantly less
than the higher one.
The most confidence level
is for spin of $J \geq 5/2$.

## The results are presented in Table I.

| $M_{exp} \pm \Delta M_{exp}$ GeV/$c^2$ | $G_{exp} \pm \Delta G_{exp}$ GeV/$c^2$ | $G_R \pm \Delta G_R$ GeV/$c^2$ | $J_{exp}$ | S.D. |
|---|---|---|---|---|
| 1.447 ± 0.007 | 0.005 ± 0.004 | 0.004 ± 0.004 | | 3.2 |
| 1.467 ± 0.003 | 0.008 ± 0.003 | 0.008 ± 0.004 | | 2.3 |
| 1.477 ± 0.002 | 0.005 ± 0.003 | $0.002^{+0.006}_{-0.002}$ | 1/2 | 3.0 |
| 1.505 ± 0.004 | 0.008 ± 0.003 | 0.005 ± 0.005 | 3/2 | 3.5 |
| 1.541 ± 0.004 | 0.011 ± 0.003 | 0.008 ± 0.004 | 5/2 | 6.8 |
| 1.606 ± 0.005 | 0.014 ± 0.005 | 0.011 ± 0.006 | 7/2 | 5.2 |
| 1.638 ± 0.005 | 0.016 ± 0.011 | $0.012^{+0.015}_{-0.012}$ | 7/2 | 3.6 |
| 1.687 ± 0.007 | 0.027 ± 0.007 | 0.024 ± 0.008 | 9/2 | 6.8 |
| 1.781 ± 0.008 | 0.029 ± 0.012 | 0.023 ± 0.015 | | 4.1 |
| 1.870 ± 0.019 | 0.036 ± 0.010 | 0.032 ± 0.011 | | 5.9 |

**The first column** contains the experimental values of the resonance masses and their errors.

**The second column** contains the experimental values of the total width of the resonances.

**The third column** contains the true widths of the resonances and their errors.

**The fourth column** contains the values of the spins of resonances (the lower limits of spins).

**The fifth column** contains the statistical significances of the resonances determined as the ratio of the number of events in the resonance to the square root of the number of background events under the resonance curve.

The estimation of the cross-section for the resonance at the mass of $M = 1.541$ GeV/$c^2$ in the $nK^+$-system from the reaction $np \to npK^+K^-$ is $\sigma = (3.5 \pm 0.7) \mu b$ at $P_n = (5.20 \pm 0.12)$ GeV/$c$.

The true width of a resonance is obtained by a quadratic subtraction of the value of mass resolution from the experimental value of the width. The function of the mass resolution increase with the increasing of masses.

For example, the value of the mass resolution is equal to ≈7 $MeV/c^2$ for the resonance at the mass of $M = 1.541$ GeV/$c^2$.

**We have tried to systematize the obtained results using the formula for the rotational bands suggested in papers of Diakonov et al. [1, 2]:** $M_J = M_0 + kJ(J+1)$ **(1),**

where: $M_J$ – the mass of the resonance, $J$ – its spin, $M_0$ – rest mass of the soliton, k – the value equal to the inversed multiplied by twice inertia moment of soliton (we use the terminology of the paper [2])

When looking to the plots of the effective mass distribution of $nK^+$-combinations one can observe that strong peculiarities are accompanied by more weak one: the weak peculiarity at the mass of $M = 1.467 GeV/c^2$, the bump at the mass region of $M = 1.565 GeV/c^2$ and others.

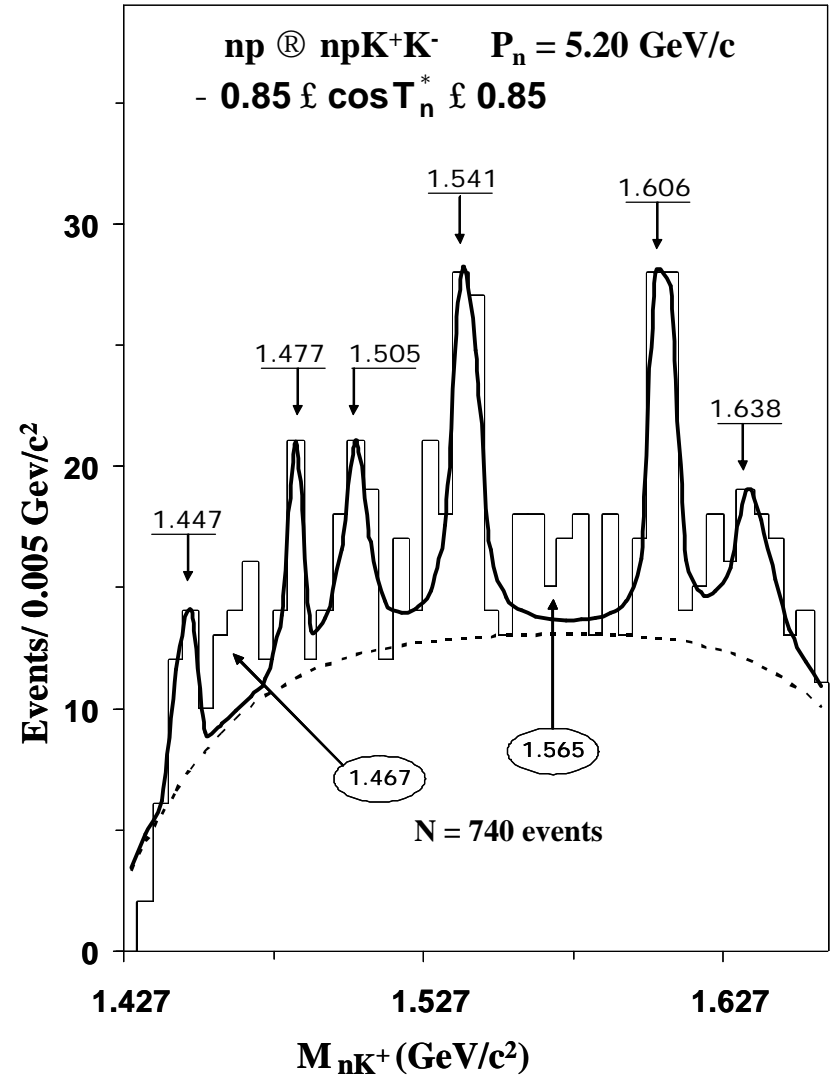

**We have tried to systematize the obtained results using the formula for the rotational bands suggested in papers of Diakonov et al. [1, 2]:** $M_J = M_0 + kJ(J+1)$       (1),

where: $M_J$ – the mass of the resonance, $J$ – its spin, $M_0$ – rest mass of the soliton, $k$ – the value equal to the inversed multiplied by twice inertia moment of soliton (we use the terminology of the paper [2])

Therefore we have carried out the approximation of the mass distributions versus spin using two variants - for "strong" resonances and for "weak" one.

    One can see a good agreement between the experimental data and the formula (1).

a)

| $M_0 = 1.462$ $GeV/c^2$    $k = 0.0090$ | | | |
|---|---|---|---|
| $J$ | $M_J$ | $M_{exp} \pm \Delta M_{exp}$ | $J_{exp}$ |
| 1/2 | 1.469 | 1.467±0.003 | |
| 3/2 | 1.496 | 1.505±0.004 | 3/2 |
| 5/2 | 1.541 | 1.541±0.004 | 5/2 |
| 7/2 | 1.604 | 1.606±0.005 | 7/2 |
| 9/2 | 1.685 | 1.687±0.007 | 9/2 |
| 11/2 | 1.784 | 1.781±0.008 | |
| 13/2 | 1.901 | 1.870±0.019 | |

b)

| $M_0 = 1.471$ $GeV/c^2$    $k = 0.0107$ | | | |
|---|---|---|---|
| $J$ | $M_J$ | $M_{exp} \pm \Delta M_{exp}$ | $J_{exp}$ |
| 1/2 | 1.471 | 1.477±0.002 | 1/2 |
| 3/2 | 1.511 | 1.505±0.004 | 3/2 |
| 5/2 | 1.565 | bump | |
| 7/2 | 1.640 | 1.638±0.005 | 7/2 |
| 9/2 | 1.736 | | |
| 11/2 | 1.854 | 1.870±0.019 | |

In Tab. (b) there are absent data about experimental values of masses and spins in third and fifth lines. There are only bumps at these masses that are not provided statistically as resonances.

In Tab. (a) the largest predicted mass at $1.901 \text{GeV}/c^2$ $(J = 13/2)$ can be cut by the phase space to the right and be observed experimentally at a lower mass.

We have done another approximation of the observed rotational bands proposing that the mass of an exited state depends not on of a resonance spin but on its orbital moment l: $M_l = M_0 + kl(l+1)$     (2).

The values of orbital moments are taken arbitrarily but so that they do not contradict the estimations of the spins. Such approximations better satisfy to experimental data.

In this description, it is necessary to take into account the resonance at $M = 1.447 GeV/c^2$.

a) – for "strong" resonances        b) – for "weak" resonances

| $M_0 = 1.481\ GeV/c^2$ | $k = 0.0100$ | |
|---|---|---|
| $l$ | $M_l$ | $M_{exp} \pm DM_{exp}$ |
| 0 | 1.481 | $1.477 \pm 0.002$ |
| **1** | **1.501** | **$1.505 \pm 0.004$** |
| 2 | 1.541 | $1.541 \pm 0.004$ |
| 3 | 1.601 | $1.606 \pm 0.005$ |
| 4 | 1.681 | $1.687 \pm 0.007$ |
| 5 | 1.781 | $1.781 \pm 0.008$ |
| 6 | 1.901 | $1.870 \pm 0.019$ |

| $M_0 = 1.447\ GeV/c^2$ | $k = 0.0100$ | |
|---|---|---|
| $l$ | $M_l$ | $M_{exp} \pm DM_{exp}$ |
| 0 | 1.447 | $1.447 \pm 0.007$ |
| 1 | 1.467 | $1.467 \pm 0.003$ |
| **2** | **1.507** | **$1.505 \pm 0.004$** |
| 3 | 1.567 | bump |
| 4 | 1.647 | $1.638 \pm 0.005$ |
| 5 | 1.747 | |
| 6 | 1.867 | $1.870 \pm 0.019$ |

Taking into account the assumption about the orbital moments, the parity of the resonance at the mass of $M = 1.541 GeV/c^2$ is negative. One can conclude taking into account in addition the value of its spin $J = 5/2$ that this resonance is not placed at the vertex of anti-decuplet suggested in papers [1,2]. But there is a probability that the resonance at the mass of $M = 1.501 GeV/c^2$ (with positive parity and spin equal to 1/2) is placed at the vertex.

Our determination of the spin for the mass of $M \approx 1.505 GeV/c^2$ does not contradict to the fact that there can be placed 2 resonances at the mass of $M = 1.501 GeV/c^2$ ($J^P = 1/2^+$) and at the mass of $M = 1.507 GeV/c^2$ ($J^P = 3/2^-$). In this case both of them are very narrow and are shifted relative to each other that gives in result the average value of an experimental mass equal to $M = 1.505 GeV/c^2$.

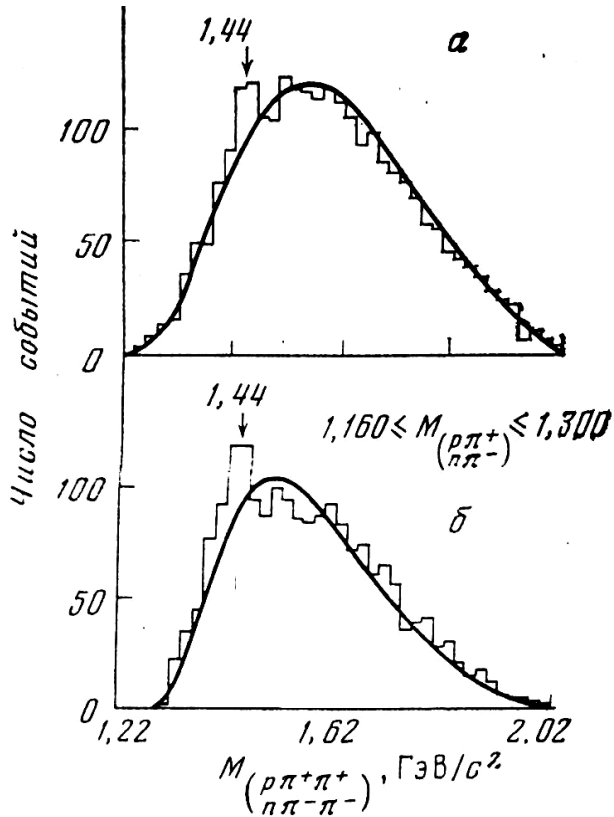

Рис. 1. Распределение эффективных масс $p\pi^+\pi^+$ ($n\pi^-\pi^-$)-комбинаций при $p_n$=5,1 ГэВ/c. $a$ — вся статистика, $б$ — из событий, где масса $p\pi^+$ ($n\pi^-$)-комбинаций заключена в пределах 1160$\leq$ $\leq M_{p\pi^+(n\pi^-)}\leq$1300 МэВ/c².

**It is necessary to do an additional remark.**

**The problem of pentaquarks arised as early as 60[th] of last century.**

**Our first studies concerning problem of pentaquarks have stimulated the realization of the unique neutron beam for 1-m HBC of LHE JINR due to acceleration of deuterons in LHE synchrophasotron.**

**In 1979 there was published our paper, about observation of rather narrow $\left(G{=}43\text{MeV/c}^2\right)$ resonance in the effective masses of $\mathbf{D^{++}p^+}$ $\left(\mathbf{D^-p^-}\right)$-combinations at the mass of M=1.440GeV/c² with statistical significance equal to 5.5 S.D. These resonances could be interpreted as five-quark states-$uuuud$ ($ddddu$) for $\mathbf{D^{++}p^+}$ $\left(\mathbf{D^-p^-}\right)$ (picture).**

**The existence of these new resonances with $J{=}I$ was predicted in papers of Grigorian and Kaidalov. Their predictions have coincided with our data.**

**We have published in 1983 the following paper about this problem using the increased statistics. There were additionally observed 2 states at the masses of M=1.522GeV/c² and M=1.894GeV/c².**

By this means the question about states containing more than 3 quarks is discussed for a long time and there are theories predicting them.